\documentclass[runningheads]{cl2emult}

\usepackage{makeidx}  % allows index generation
\usepackage{graphicx} % standard LaTeX graphics tool
\usepackage{subeqnar} % subnumbers individual equations
\usepackage{multicol} % used for the two-column index
\usepackage{cropmark} % cropmarks for pages without
\makeindex            % used for the subject index
\begin{document}
\title*{Population III by Popular Demand\protect\newline
 -- Progress and Previews}
\toctitle{Population III by Popular Demand -- Progress and Previews}
% allows explicit linebreak for the table of content
%
%
\titlerunning{Population III by Popular Demand}
% allows abbreviation of title, if the full title is too long
% to fit in the running head
%
\author{Timothy C. Beers}
\authorrunning{Timothy C. Beers}
% if there are more than two authors,
% please abbreviate author list for running head
%
%
\institute{Department of Physics \& Astronomy, Michigan State University,\\
E. Lansing, MI 48824  USA}

\maketitle              % typesets the title of the contribution

\begin{abstract}

I discuss the ongoing search for stars of the Milky Way  which have been
referred to as members of Population III.  Following a discussion of possible
definitions for these stars, I consider the reasons why astronomers have
undertaken this search, and list some of the numerous astrophysical uses of the
extremely metal-poor stars which have been found along the way.  I then
review survey techniques which have been used in the past, and provide an
update on plans for future investigations.  Finally, the question of
when one might consider the search for Population III FINISHED is addressed.

\end{abstract}

\section{Introduction}

The most metal-deficient stars in the Milky Way provide astronomers with a
wealth of information on the early epochs of galaxy formation, the nature and
interplay of various stellar populations, and even on the chemical composition
of the (nearly) metal-free Universe shortly after the Big Bang.  Since the
topic of this meeting is the FIRST STARS to have formed in the Universe, in
this brief review I consider several fundamental questions which must be
addressed in order to make progress in our understanding of these objects: (1)
What is a Population III star ?, (2) Why search for Population III stars ?, (3)
How should we search for Population III stars ?, and finally, (4)
When do we stop the search for Population III Stars ?

\section{What IS a Population III Star ?}

Population III is a term which has taken on a variety of meanings in the past,
as astronomers have refined their picture of early star formation in the
Universe.  Population III means different things to different people !  Among
the definitions which have been employed:

\begin{itemize}
\item
A star with metal abundance which is significantly less than the lowest
abundance globular clusters ([Fe/H] $\sim -2.5$)
\item
A star of ``pure'' Big Bang composition -- H, He, Li, and
precisely ZERO heavier elements
\item
A star which was born within the first $10^6-10^7$ years following the Big
Bang (independent of its metal abundance)
\item
A star which is no longer ``living,'' but which was formed very early in the
history of the Universe and {\it imprinted} telltale signs of
its previous existence in the elemental abundance patterns of presently
observable stars
\end{itemize}

If we take this last definition for a Population III star, it must be
acknowledged that we will never be able to point our telescopes to a given
position on the sky and directly examine the properties of the first stars to
have formed in the Galaxy.  For the theoretician, this is not particularly
troublesome.  For the observer, this is less than satisfactory.  Since there is
not a single star known to date which is widely agreed upon as being a member
of Population III, we still have the luxury of choosing a working definition of
what these objects might be.  Exercising the unlimited optimism of the
observational astronomer, I propose the following working definition for a {\it
detectable} Population III star:
\par
\bigskip
\noindent {\bf A Population III star is a presently-surviving (i.e., $\tau_{\rm
ms} \ge 14$ Gyr) member of the very FIRST generation of stars born in the
Galaxy, from gas of primordial abundance.  In other words, a truly ZERO metal
abundance star (for all intents and purposes, a star with a measurable
atmospheric abundance [Fe/H]$ \le -6.0$).}
\par
\bigskip
Note that this definition implies a very specific set of requirements:

\begin{itemize}
\item
The FMF (First Mass Function) must extend to include low mass stars with
sufficiently long main-sequence lifetimes that survivors might be identifiable
at present
\item
Pollution of a stellar atmosphere (either by intrinsic or extrinsic
mechanisms) with metallic species subsequent to a star's formation cannot be
effective (else we would never measure a star's atmospheric metal abundance
satisfying the definition)
\end{itemize}

This definition is useful because it is testable.  If, over the course of the
extensive searches described below, astronomers still find themselves without a
single Population III star to point their telescopes at, then we will have
learned that one or both of the above requirements must not have been met in
nature.

\section{Why Search for Population III Stars ?}

Unfortunately, we cannot yet answer this question with the old standby ``because
they are there,'' as we don't yet know that !  However, even a completely
unsuccessful search for Population III stars rewards astronomers with insight
into a wide variety of astrophysical issues.  From spectroscopic and
photometric information gleaned from the ``rejects'' of searches for
Population III stars one can extract nuggets of precious information.  For
example: 

\begin{itemize}
\item
The nature of the halo Metallicity Distribution Function (MDF) -- how low can
we go ?
\item
Abundances of heavy metal species in the SECOND generation of stars
\item
Measurements of the primordial lithium abundance and limits on the flux of
early Galactic cosmic rays
\item
Direct measurements of the elemental yields of early Type II supernovae, and
the opportunity to unravel their mass distribution
\item
Identification of the astrophysical site(s) of the $s$-process(es)
\item
The nucleosynthesis products of the first AGB stars
\item
Identification of the astrophysical site(s) of the $r$-process(es?)
\item
Efficiency of mixing processes in the early Galaxy
\item
Measurement of elemental species (such as Zn) which are used to explore gas
abundances along lines of sight toward distant QSOs
\item
Estimation of hard lower limits on the age of the Galaxy and the Universe from
the use of light- and heavy-element chronometers
\item
Evidence for the existence and nature of the metal-weak tail of the
thick disk of the Galaxy
\item
Measures of the halo and thick disk velocity ellipsoids, and their alteration
with Galactocentric distance
\item
Examination of kinematic constraints on the so-called ``counter-rotating high
halo'' and other evidence, such as kinematic substructure, of past and present
mergers of the Galaxy with smaller satellites
\end{itemize}

Many of the above issues are discussed in this volume.  In this review I
discuss the stars which form the observational database upon which such
investigations are based.

\section{How Should We Search for Population III Stars ?}

Ever since astronomers realized that extremely metal-poor stars are rare, but
identifiable, members of the presently observable Milky Way, we have been
working hard to devise effective survey strategies to ferret them
out in ever increasing numbers.  Quite independent of the techniques applied,
there emerge two fundamental ``barriers'' which must be overcome in order for a
search to be successful: 

\begin{itemize}
\item
A successful survey technique must maximize the ``effective yield'' (EY) of
interesting candidates, so that large amounts of telescope time are not wasted
obtaining follow-up observations of stars which turn out not to be of interest
\item
A successful survey technique must be applied to a large fraction of (at least)
the high Galactic latitude sky, so as to obtain a significant increase in the
numbers of recognized extremely metal-poor stars
\end{itemize}

For the purposes of this discussion, let us define the effective yield of a
given survey as:

\begin{equation}
{\rm EY} = \frac{[\#\; {\rm stars \; with\; [Fe/H]} \le -2.0]}{[{\rm total\; \# \;of\; targeted\; stars}]}
\end{equation}

\par
For an efficient follow-up campaign, an ``ideal'' EY would be higher than 50\%.
This has not yet been achieved, but should be possible with concentrated
effort.  Past experience suggests that it is often difficult to obtain
sufficient telescope time for the demonstration and refinement of a new survey
technique (especially when one is confined to shared, rather than private,
observatories).  The end result is that compromises are made in the EY of the
survey.  This is a most unfortunate reality.

The ``sky coverage barrier'' has resulted in surveys which require many
years, if not decades, to come to fruition.  This is a particularly acute
problem when one is relying on shared facilities.  My involvement with the HK
survey, for example, began in 1983 (as an exercise in learning a spectrographic
data reduction software package in those pre-IRAF/MIDAS days we would all like
to forget).  Over the past (gasp) 16 years, the follow-up efforts from the HK
survey have expanded to involve over 30 astronomers working with $\sim 15$
different telescope/detector combinations, with the final total number of
(medium-resolution) spectroscopic observations reaching roughly 5000 stars.  I
am extremely grateful for the patience of those collaborators who have ``stayed
the course'' throughout this long journey.  We have succeeded in identifying
over 100 stars with [Fe/H]$\; < -3.0$, a few stars near [Fe/H]$\; = -4.0$, and
of course, many other interesting stars of somewhat higher abundance.

Within the next 2-3 years, ALL of the presently-recognized stars with
abundances [Fe/H] $\;< -3.0$ will be observed at high resolution (and
presumably with exquisite signal-to-noise ratios) with new-generation
telescopes.  This is a sobering statement !  I address what might be done about
this state of affairs in a separate contribution to this volume.

\subsection{Searches in the Past}

Before we consider new strategies in the search for Population III stars, it is
instructive to consider the approaches which have been employed in the past.
Space precludes a complete discussion, but a list of previously-applied survey
techniques would include the following.

\subsubsection*{``Informed Serendipity.''}

This term (first used in this context by John Norris), refers to the targeting
of a small number of candidate stars with characteristics which are suggestive
of the most extreme metal-poor stars.  For example, the discovery that G64-12
is a main-sequence dwarf with abundance [Fe/H]$\; \sim -3.5$ came 
an investigation of stars with the largest known space
motions \cite{cp81}.  The most metal-deficient giant yet known, CD$-$38:245,
with [Fe/H] $\sim -4.0$ \cite{bn84}, came from a limited follow-up
of stars identified from the Slettebak \& Brundage prism survey \cite{sb71},
and noted by them as having ``a continuous spectrum'' but shown by
\cite{gs73} to have colors typical of a rather late-type star.  In spite of
great observational effort since then, there are presently only a handful of
stars known with metallicities at or below [Fe/H] $= -3.5$.  ``Informed
Serendipity'' is a powerful technique.

\subsubsection*{High Proper Motion Samples.}

Spectroscopic surveys of stars selected to have high proper motions have
provided roughly 10\% of the known stars with [Fe/H] $\le -2.0$.  Examples of
such surveys include the lists of Luyten proper motion stars in the southern
hemisphere studied by Ryan \& Norris \cite{rn91}, and the Giclas proper motion
sample in the northern hemisphere investigated by Carney et al. \cite{cea94}.
Unfortunately, the EY of these surveys is not particularly high.

\subsubsection*{Wide-Field Objective-Prism Surveys.}

Photographic objective-prism surveys over large areas of sky, based on widened
spectra and a visual selection technique, have provided the majority of the
presently known extremely metal-poor stars.  One example is the survey of Bond
(\cite{b70},\cite{b80}), which resulted in the identification of samples of
metal-poor stars that continue to be discussed in the modern literature, but
which only contained a relatively small number of stars with [Fe/H] at or
below $-2.5$, leading Bond to ask ``Where Is Population III ?'' \cite{b81}.

George Preston suggested a slightly modified version of this approach,
employing an interference filter in the focal plane, used in combination with
the objective prism.  This effort, now known as the HK survey (formerly
referred to in the literature as the Preston-Shectman survey), was able to
go several magnitudes deeper than the Bond survey, and has been described in
detail in \cite{bps85},\cite{bps92}, and \cite{b99}.  This technique has been
successful in the identification of roughly 1000 halo stars with
[Fe/H] $< -2.0$, and has provided the majority of targets of immediate interest
for high-resolution spectroscopic study with the new-generation 8m-class
telescopes (such as VLT, SUBARU, and GEMINI) which are now coming online.

Based on numbers presented in Table 1 of \cite{b99}, the EY of the HK survey
is on the order of 22\%, so roughly twice that of proper-motion selected
samples, but still at least a factor of two smaller than the ideal case.  The
careful reader of this table will note a rather wide variation in EY among the
groups which carried out the spectroscopic follow-up campaigns.  This variation
arises, in the main, because not all groups were in a position to conduct a
photometric pre-selection of the original HK survey candidates prior to
obtaining spectroscopy, a filtering process which can effectively eliminate
stars which are either too hot or too cool to be of interest.  When a
photometric pre-filtering {\it was} done (e.g., the BPS and ANU groups), 
the HK survey achieved an EY of $\sim 32\%$.

We are presently obtaining digital scans of the HK survey plates
using the APM facility in Cambridge.  Color calibrations of direct sky survey
plates in the northern and southern hemisphere are expected to
be completed shortly as well.  With this data in hand, we should be able to
recover a large fraction of the cooler metal-deficient stars on the HK survey
plates which were missed during the visual selection procedure.  Preliminary
results of an artificial neural network analysis of this data has been
presented in \cite{rea99}.  All indications are that the EY resulting
from this procedure should reach close to 50\%.

\subsection{Searches in the Future}

How can we do better ?  There are a number of surveys which are just getting
underway that hold the promise of dramatically increasing the numbers of
known extremely metal-poor stars, and, perhaps, will result in the discovery of
a {\it bona-fide} Population III star.

\subsubsection*{The Hamburg/ESO Survey.}

The Hamburg/ESO Stellar Survey (HES) described by Christlieb et al. (this
volume, and references therein) is based on automated scans of {\it unwidened}
wide-field objective-prism plates, covers essentially the entire southern
Galactic cap, and reaches a limiting magnitude of $B \sim 17-17.5$, 1-1.5
magnitudes fainter than that reached by the HK survey.  Since each one
magnitude interval in limiting magnitude results in an increase in survey
volume of a factor of 4, one might expect the HES to eventually identify on the
order of $\sim 5-10$ times as many extremely metal-poor stars as the HK survey.
Of course, eventually one reaches distances where the rapidly falling density
profile of the stellar component of the halo limits the increase in the numbers
of interesting stars to be detected at fainter apparent magnitudes.  The real
advantage of the HES over the HK survey is due to the quality of the prism
spectra, their greater wavelength coverage, and the automated selection
criteria.  These advantages assure that low-metallicity candidates can be
selected without the introduction of a metallicity-temperature bias, a further
limitation of the (visual) HK survey.  As a result, the expected EY of
interesting metal-poor stars from the HES survey should be substantially higher
than that of the HK survey.  Tests of the EY are presently being conducted.

\subsubsection*{Broad-Band Colorimetric Surveys.}

Since we believe that Population III stars (according to the above definition)
are most likely to be found in the halo of the Galaxy, it makes sense to
conduct surveys for possible candidates {\it in situ}.  One efficient
way to do this is by a careful pre-selection of stars which are, for example,
main-sequence dwarfs near the tip of the halo turnoff, and that exhibit colors
consistent with a complete (or nearly complete) absence of heavy
metals in their atmospheres.  Accurate broad-band photometry is now becoming a
reality for literally hundreds of millions of stars with the advent of the
Sloan Digital Sky Survey (SDSS).  As described in \cite{lea98}, properly
combined filter observations of faint halo stars should be able to readily
separate stars with abundance [Fe/H]$\;\le -2.0$ from those above this value.
Although there is expected to be little sensitivity to metallicity below --3.0,
the large lists of candidate metal-poor stars to be generated from SDSS will
provide very attractive targets for future study.  The EY of the SDSS lists of
metal-poor stars has yet to be determined, and tests of the variation of the
EY with apparent magnitude (due to the greater photometric errors for the
fainter stars, and to the potential intrusion of large numbers of fainter stars
of normal metal abundance into the sample) have yet to be carried out, but all
present indications at present are encouraging.

Low mass M-type stars potentially comprise the {\it dominant} stellar component
of the Milky Way halo, yet only recently has their nature come under scrutiny.
Because of their great numbers, and under the assumption that the FMF does not
change significantly at low abundances, the sdM stars provide excellent samples
for examination of subtle variations in the halo MDF as a function of distance,
and in the search for Population III.  At present only a handful of sdMs are
known with [Fe/H] $\le -1.0$, mostly from spectroscopic examination of high
proper-motion catalogs \cite{gr96}.  However, Gizis \cite{g97} has shown
that sdMs can be readily recognized from inspection of various two-color planes
(e.g., $J-H \; vs.\; H-K$), and further, that reliable estimates of abundance
can be obtained from moderate-resolution spectroscopic analysis of their
molecular bands (CaH, TiO) in the spectral region $6300 \le \lambda \le 7200$
\AA\ . It has been estimated that on the order of 15,000 sdMs
with [Fe/H] $\le -1.0$ can be identified from follow-up spectroscopy of
candidates selected from the recently-initiated 2Mass survey \cite{g98}.  We
await the initiation of a dedicated spectroscopic follow-up campaign in order
to test the EY of this approach.

\subsubsection*{Narrow-Band Colorimetric Surveys.}

McWilliam and collaborators have begun exploring the use of a narrow-band
filter, centered on the stellar CaII H and K lines to obtain CCD photometry of
stars, and thereby select extremely metal-poor stars by the lack of strength in
these resonance lines \cite{m98}.  The first application of this
technique to stars in the Galactic bulge and in nearby dwarf galaxies appears
quite promising.  Again, the crucial issue will be the demonstration of a
useful EY for this approach.

\subsubsection*{Wide-Field Spectroscopic Surveys.}

Single-slit spectroscopic follow-up of the large numbers of newly-identified
candidate extremely metal-poor stars from the surveys described above would be
a daunting observational task.  Fortunately, a far more efficient follow-up
technique is now feasible.  The pioneering work of Watson and Parker has given
the astronomical community the FLAIR II wide-field spectrograph, now in
operation at the UK Schmidt in Australia.  Recent approval of the AAT Board for
preparation of a robotic fiber positioning system for the FLAIR II instrument
will result in a new six-degree field (6DF) instrument to be in place in late
1999 \cite{w98}.  By early 2000, it is anticipated that the 6DF will be in
routine operation.  With the benefit of a CCD upgrade, and some 150 active
fibers, acceptable integration times will yield $\sim 2$ \AA\ resolution
spectra of stellar sources down to $B = 16.5$.  One interesting challenge
remains:  filling all the fibers on the 6DF with useful target stars !
Clearly, the optimal approach is to combine catalogues extracted from projects
with similar wavelength and resolution requirements.  The large numbers of
stars to come from the digital HK survey, the HES, the southern-most (and
brighter) stars from the SDSS, as well as additional proper-motion selected
stars, suggest themselves.  The sdM stars, owing to their lack of flux in the
blue, would probably have to be undertaken as a separate survey.  Obviously,
developing a similar instrument to the 6DF to be used on northern-hemisphere
Schmidt telescopes should be given high priority as well.

There is little doubt that the search for Population III stars will reach
a pinnacle with the successful operation of planned space-based survey
instruments such as GAIA \cite{gi98}.  The possibility of obtaining
medium-resolution spectroscopy (as well as proper motions, and accurate
astrometry) for up to $10^7$ stars down to 17th magnitude will revolutionize
Galactic astronomy in virtually every way.  Assuming a successful mission, one
might predict an EY approaching 100\%.

\subsection{Are we Looking in the Right Place ?}

The results of the HK survey indicate that the lowest abundance
stars which are presently to be found in the Galaxy terminate at around [Fe/H]
$= -4.0$, well above the metallicity we would associate with true Population
III \cite{b99}.  However, it must be kept in mind that the HK survey only
identifies halo stars no greater than 5-10 kpc from the Sun (for giants; 1
kpc is the approximate limit for main-sequence turnoff stars in the halo).  It
may well be the case that such locally-selected samples of stars are not
representative subsets of the most metal-deficient stars to be found in the
Galaxy.  One additional complication has now become clear --  the thick disk of
the Galaxy (which has a space density up to 50 times that of the halo component
in the solar neighborhood) does indeed possess a metal-weak tail (see Chiba,
this volume, and references therein), at least down to [Fe/H] $\sim -2.0$.  The
effect of these local ``contaminants'' from the thick disk may
artificially boost the apparent numbers of metal-poor stars in the solar
neighborhood to such a degree that extrapolations of the MDF based on local
samples may in fact be using an incorrect ``normalization,'' and we may have to
contend with problems of interpretation.

Tests of whether or not the local volume is a fair subsample will only come
when samples of much fainter stars, located 10's of kpc away, become available.
Some information is potentially available from the existing samples, based
on their kinematics.  One could, for example, test if the MDF of stars on
orbits which take them well outside the solar neighborhood differs greatly from
that of stars which are confined to several kpc away.  As some have speculated
\cite{r98}, if the epoch of ultra metal-poor star formation precedes the
dissipative formation of galaxies (as is likely if early generation stars are
the sources of the ionizing background radiation at $z > 5$), then the most
metal-deficient stars may fact be distributed rather similarly to the dark
matter itself, i.e., with a much broader spatial distribution than the majority
of luminous halo stars.  If this is the case, searches for Population
III will only succeed when they probe the VERY faint stars in the outskirts of
the Galaxy.

\subsection{Are we Looking for the Right Thing ?}

Thus far, our discussion has been predicated on the assumption that the surface
abundances of heavy elements in low-mass halo stars do not change over time
(the second of our requirements for a ``detectable'' Population III star).
There is evidence now accumulating to suggest that this may in fact {\it not}
be true.

\subsection{Intrinsic Pollution}

New stellar evolution models for extremely metal-poor stars \cite{fea99}
suggest that enhanced mixing at the time of helium-core flash may result in the
conversion of some quite metal-deficient stars into carbon-enhanced stars, a
mechanism which might explain the fact that $\sim$25\% of the stars in the HK
survey with [Fe/H] $\le -3.0$ exhibit strong C and N features in their spectra
\cite{ros99}.  The tantalizing possibility is that ALL low mass
stars with [Fe/H] $\le -4.0$ undergo this conversion, which means that in order
to find the stars of Population III, we should be searching for the most
metal-deficient CARBON stars in the Galaxy.  The HES discussed above is able to
preferentially select such stars based on the presence of strong CH G-bands in
their spectra, hence a dedicated follow-up of these candidates may prove quite
interesting.

\subsection{Extrinsic Pollution}

Other mechanisms by which stars which were born with [Fe/H] $ \le -6.0$ might
be missed in present-day searches for Population III have been discussed
in the literature. One idea (\cite{y81},\cite{i83}) is that, over their
extended lifetimes, Population III stars which repeatedly pass through the
gradually more metal-enriched ISM of the Galactic disk could accrete sufficient
amounts of heavier metals onto their atmospheres to boost their observed
atmospheric abundances to [Fe/H] $> -4.0$.  We now have sufficiently large
samples of metal-poor stars to test such a hypothesis.  For example, one
might compare the MDF of main-sequence dwarfs (which to first order should
retain evidence of the accreted material on their surfaces) with that of
evolved giants (which should mix the accreted material because of their
convective envelopes) -- the expectation being that the dwarf MDF should be
truncated and lack a tail at lower metallicity when compared to the giant MDF.
Another test would be to compare the abundance distribution of main-sequence
dwarfs which populate ``plunging'' orbits, and hence spend a tiny fraction of
their time passing through the Galactic disk (and do so at high velocity) with
that of similar stars which spend their lifetimes on orbits which
continually bathe them with disk ISM, hence maximizing the possibility for
accretion.  Large catalogs of metal-poor stars with available space motions
\cite{bea99} should provide useful samples for such an investigation.

\section{When Do We Stop the Search for Population III Stars ?}

Now seems an appropriate time to consider the longer-term prospect for
conducting searches for evidence of the elusive Population III.  Below I
consider four possible ``termination points.'' 

\subsection{Now ?}

I am asked all of the time whether or not we have identified a sufficiently
large number of extremely metal-poor stars already.  The answer I give is
always the same, a resounding ``NO !''.  The samples we have assembled over the
past two decades have certainly opened investigations which may in
fact resolve a number of the long-standing questions of early Galactic
nucleosynthesis.  However, as soon as one moves beyond the most obvious
questions, and begins to subdivide the existing samples of extremely metal-poor
stars based on, for example, (1) their kinematics and orbital families, (2)
their light- and heavy-element abundance patterns, or (3) their present
evolutionary states, it becomes obvious that the individual subsamples are
going to be far too small to obtain adequate descriptions of general patterns.

As mentioned above, the blue-sensitive, high resolving-power
spectrographs such as UVES on the VLT-UT2, HDS on SUBARU, and HROS on GEMINI,
will shortly provide astronomers with high-resolution data for the entire
sample of Galactic stars which are known at [Fe/H] $< -3.0$.  If, at the end of
this time, we expect to have unanswered questions about the nature of the early
Galaxy as revealed by observations of the most metal-deficient stars (and who
does {\it not} expect this ?), there will exist a need for at least another
several hundred stars with [Fe/H]$ \le -3.0$.  Given the minimum several-year
lead time which is required to assemble such samples, now is the time to
re-double our efforts.
 
\subsection{In the Near Future, $\sim$5--10 years ?}

The next decade will be a transition period for studies the first stellar
populations of the Galaxy.  Large surveys of the type I have described above
should be brought to completion by this time.  This will not happen
automatically, of course.  As a community, it is important that we utilize the
survey telescopes of the future -- the 4m class telescopes of today -- to
obtain the medium-resolution spectroscopy and photometry which is required to
fully exploit the dedicated searches for Population III.  We encourage 
telescope time allocation committees to consider the ``fast rather than faint''
model for assigning access to those who wish to pursue the long-term follow-up
observations which will be needed to make significant progress.

One can safely predict an explosion of detailed understanding of the elemental
abundances of first- and second-generation stars from the data flow generated
by 8m-10m class telescopes.  The era of space-based survey missions will be
upon us, and ground-based surveys will have less and less impact on the
continued searches for Population III, at least for stars of apparent magnitude
$<$ 17.

\subsection{In the Distant Future, $\sim$10--15 years ?}

GAIA, and other missions (such as SIM) will be well underway, and perhaps
completed.  One might speculate that astronomers, by this time,
should have intimate knowledge of the nature of the early Galaxy, and Universe.
Perhaps this will be so.  Experience suggests, however, that the deluge of new
information acquired will show us the ``mistakes of the past'' and suggest
interesting new avenues for future observational campaigns.

\subsection{Never} 

Never is a long time, but I suspect that the search for
Population III, and the knowledge obtained about the nature of the extremely
metal-poor stars of the Milky Way, will continue to amuse, delight, perplex,
and hopefully, enlighten astronomers for a long time to come.  This is where
the smart money is going.

%INDEX%%%%%%%%%%%%%%%%%%%%%%%%%%%%%%%%%%%%%%%%%%%%%%%%%%%%%%%%%%%%%%%
\clearpage
\addcontentsline{toc}{section}{Index}
\flushbottom
\printindex
%%%%%%%%%%%%%%%%%%%%%%%%%%%%%%%%%%%%%%%%%%%%%%%%%%%%%%%%%%%%%%%%%%%%%

\end{document}